\documentclass[a4paper,11pt]{article}

\usepackage{amsmath}
\usepackage{amssymb}
\usepackage{amsfonts,latexsym}
\usepackage{graphicx}
\usepackage[authoryear]{natbib}

\usepackage{lineno}

\title{Parameter variability can produce heavy tails in a model for the spatial distribution of settling organisms}
\date{}

\author{Luis F. Gordillo\\ 
{\small Department of Mathematics and Statistics, Utah State University, Logan, UT}\\
Priscilla E. Greenwood\\ 
{\small Department of Mathematics, University of British Columbia, Vancouver, BC}}

\begin{document}
\maketitle
\begin{abstract}
We show that a simple mechanistic model of spatial dispersal for settling organisms, subject to parameter variability, can generate heavy-tailed radial probability density functions. The movement of organisms in the model consists of a two-dimensional diffusion that ceases after a random time, where the parameters that characterize each of these stages have been randomized. Our findings show that these minimal assumptions can yield heavy-tailed dispersal patterns, providing a simplified framework that increases the understanding of long-distance dispersal events in movement ecology.
\end{abstract}

\section{Introduction} 
The mechanisms underlying long-distance dispersal in ecology remain partially understood despite the significant theoretical advances and extensive data collection over the past decades (\cite{Bullock, Clark99, Fandos2023}). In part, this is due to the intricate interplay of various factors, including behavioral, individual and environmental complexities, which vary between scenarios (\cite{Morales22, Nathan11, Schupp19}).
As a result, dispersal kernels (or equivalently, density functions of locations) often oversimplify the underlying processes and sources of variability (\cite{Bullock, Nathan}). In the case of long-distance dispersal, kernels are frequently fitted with limited regard for the underlying biological mechanisms driving these patterns (\cite{Bullock, Clark05, Snell_et_al_2019}).

Dispersal kernels with tails that decay more slowly than that of a Gaussian have been shown to arise when diffusivity is treated as a random variable drawn from a specified distribution (\cite{Petrovskii08, Petrovskii09}). However, such analyses overlook the fact that individual organisms, referred to here as \textit{individuals}, eventually cease movement at random times, which significantly influences the shape of dispersal kernels (\cite{Morales22}). When the stopping times are very long,  individuals would be expected to diffuse further.
In this paper, we show that the probability density of individuals' final locations can have spatially heavy-tailed behavior, e.g. the tail decays as a power-law, emerging from a two-step model of movement and settlement when the distribution of settling times is heavy-tailed (in time). 
This is an alternative approach to long distance dispersal models, like L\'evy walks, which describe movement patterns with step lengths that follow a heavy-tailed distribution (\cite{Viswanathan}) or advection-diffusion models accounting for turbulence (\cite{Nathan11}).

We start by recalling the classic paper by Broadbent and Kendall (\cite{BK53}), where a model for the dispersal of larvae of helminth parasites was proposed. Our paper is based on their approach. 

Broadbent and Kendall considered the spatial movement of each individual larva as a Brownian motion in the plane, with the origin as the initial position. Individual movement stops after an exponentially distributed time, which is the time the larva climbs onto a grass blade and is eaten by a grazer. 
The Broadbent-Kendall model is conceptually straightforward; its core idea has been well studied (\cite{Lewis, Renshaw}) and adapted to contexts with similar dynamics (\cite{Williams, Yasuda}). 
The model yields the radial probability density function for the larva's final  position. This density depends on two parameters: a diffusion coefficient that governs movement and a parameter of the exponential distribution of the stopping time.

In the Broadbent-Kendall model, the tail of the spatial density decays exponentially, so that it is light-tailed, which makes the model unsuitable for describing the outcome of long distance dispersal mechanisms.
However, to introduce more flexibility to account for intrinsic or extrinsic variability, it has been found that such parameters can be represented by distributions (\cite{Clark99, Nathan, Petrovskii08, Petrovskii09}). 
For instance, seed settling rates can vary with environmental conditions, while insect dispersal may differ due to physiological traits. This prompts the question: How does the stochasticity of the parameters affect the shape of the density function of Broadbent and Kendall?
Here, we start with exponentially stopped diffusions and allow the parameters governing both diffusion and settling to follow familiar distributions. The resulting spatial tail of the dispersal distribution will have either light or heavy tail behavior depending on the distributions used for each parameter. 
We derive simplified analytical expressions for the spatial probability distributions obtained and compare their spatial tails with those derived using fixed parameters. 

This paper is organized as follows: In Section 2 we derive explicit analytical expressions for the radial probability density of individuals' final locations, after randomizing each one of the two parameters and then both. In Section 3 the tail analysis is presented. 

\textbf{Terminology.}  A \textit{light tail}, or thin tail, distribution has a tail probability that decays \textit{at least as fast} as an exponential function $e^{-\theta x}, \theta>0$. A \textit{heavy tail} distribution has a probability tail that decays more slowly than an exponential function.

\section{Randomizing the parameters}
\subsection{Random settling rate $\lambda$ }
Our assumption is that each individual moves in the plane with Brownian motion and stops at a time $\tau$, the \textit{settling time}. In (\cite{BK53}), $\tau$ is exponentially distributed with parameter $\lambda>0$, the \textit{settling rate}. 
Suppose that $\lambda$ is not the same for all individuals but has a probability distribution  within the class of individuals, i.e. $\lambda$ is a random variable with probability density $f_\lambda (s)$ having support $(0,\infty)$. The distribution of the settling time is now
\begin{equation}\label{eq: settling rate}
H(t)=P(\tau \leq t)=\int_0^\infty P(\tau \leq t|\lambda=s)f_\lambda(s)ds 
= \int_0^\infty (1-e^{-st})f_\lambda(s)ds 
= 1-\mathcal{L}(f_\lambda(s))(t),
\end{equation}
where $\mathcal{L}$ denotes the Laplace transform.

Following \cite{BK53}, we assume that each individual undergoes diffusive movement in the plane, starting from the origin, and stopped at a time with distribution (\ref{eq: settling rate}). We assume that the movement is isotropic and denote the diffusion coefficient by $\sigma^2 > 0$. Let $R(t)$ represent the distance of an individual from the origin at time $t$. In polar coordinates, the probability that $R(t)$ lies within the interval $[r, r + dr)$ is given by
\begin{equation}
\phi(r,t)dr=P(r\leq R(t)< r+dr)= \frac{2r}{B^2t}e^{-r^2/B^2t}dr,
\label{eq: probability radius}
\end{equation}
where $B^2=2\sigma^2$, to which we will refer as the \textit{diffusion parameter} or \textit{diffusivity}. By combining the expressions (\ref{eq: settling rate}) and (\ref{eq: probability radius}) we derive the radial probability density, $g(r)$, of the final location of an individual, 
\begin{equation}\label{main formula}
g(r)=\int_0^\infty \phi(r,t)\cdot \frac{dH}{dt}dt=\int_0^\infty \phi(r,t)\cdot \mathcal{L}(sf_\lambda(s))(t)dt.
\end{equation} 
In the case when $\lambda>0$ is constant and  $H(t)=1-e^{-\lambda t}$, Broadbent and Kendall identified the radial density  
\begin{equation}
    g(r)=\frac{4\lambda r}{B^2}K_0\left(\frac{2r\sqrt{\lambda}}{B}\right),
    \label{B and K density}
\end{equation}
where $K_0$ is a modified Bessel function of the second kind. As will be verified in Section 3, this density is light-tailed.

As applications of equation (\ref{main formula}), let us derive the radial density of individuals, $g(r)$, for two cases: when $\lambda$ is uniformly distributed and when it follows a Gamma distribution. 
In practice, when little prior information is available, $\lambda$ might be assigned a uniform distribution over a plausible range, chosen based on physical or biological constraints. As noted in (\cite{Clark99}), the Gamma distribution is commonly used for modeling parameter variability due to the flexibility of its shape and its conjugacy with the exponential in Bayesian statistics. Since our motivating examples show long-distance dispersal, it is necessary that the combination of the two distributions, in space and time, produce heavy tails. This can be achieved when the values of $\lambda$ follow a Gamma distribution.

\subsection{Examples with random $\lambda$} 
\subsubsection{$\lambda$ is uniformly distributed}
\begin{figure}
\includegraphics[scale=0.55]{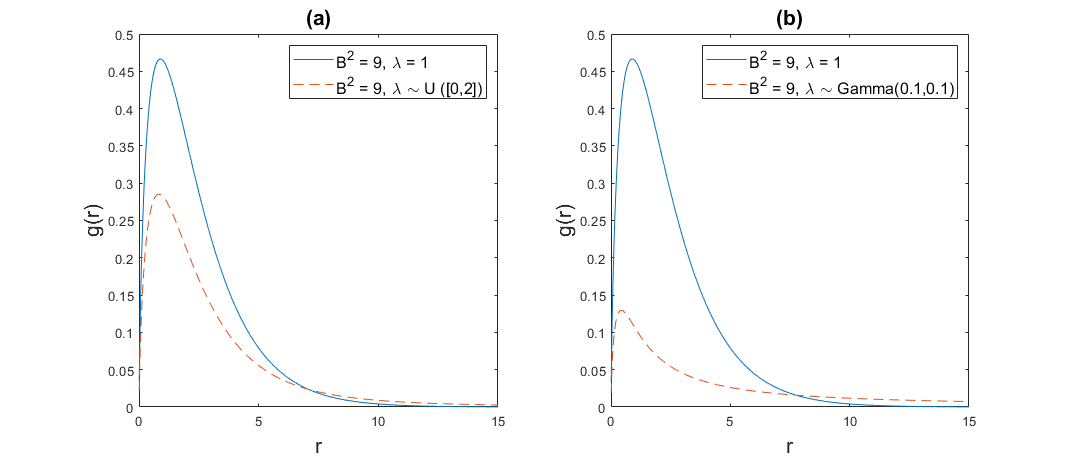}
\centering
  \caption{Probability density for the radial distribution of individuals, $g(r)$. The Broadbent-Kendall model, with $\lambda=1$ and $B^2=9$, appears as the continuous curve. For $\lambda$ (a) uniformly distributed and (b) Gramma distributed, the density appears as the dashed curve.}
\label{fig: BK lambda random}
\end{figure}

If $\lambda$ is uniformly distributed in $[a,b]$,
the probability that an individual stops moving before time $t$ is
\begin{equation}
H(t)=P(\tau \leq t)=1-\frac{e^{-at}-e^{-bt}}{(b-a)t}
\end{equation}
and the probability density for the radial distribution of individuals is obtained using (\ref{main formula}),
\begin{equation}
g(r)=\int_0^\infty \frac{2r}{B^2 t}e^{-r^2/B^2t}\cdot \frac{(at+1)e^{-at}-(bt+1)e^{-bt}}{(b-a)t^2} dt,
\end{equation}
which in terms of the modified Bessel functions of second kind $K_1$ and $K_2$ is
\begin{equation}
g(r)=\frac{4}{b-a}\left( \frac{a^{3/2}}{B}K_1\left(r\frac{2\sqrt{a}}{B} \right)+\frac{a}{r} K_2\left(r\frac{2\sqrt{a}}{B} \right) - \frac{b^{3/2}}{B}K_1\left(r\frac{2\sqrt{b}}{B} \right)-\frac{b}{r} K_2\left(r\frac{2\sqrt{b}}{B} \right)  \right).
\label{eq: lambda uniform, B constant}
\end{equation}
This expression is obtained by using a standard characterization for modified Bessel functions, see (\cite{Gradshteyn}) pg. 368 (9). 
Figure \ref{fig: BK lambda random}(a) displays the plots of the Broadbent–Kendall model in equation (\ref{B and K density}) with $\lambda = 1$ (solid curve), and the probability density in equation (\ref{eq: lambda uniform, B constant}) for $\lambda$ uniformly distributed over [0,2] (dashed curve). As expected, the probability mass in the latter case is shifted to the right. However, the plot suggests, and this can be shown analytically, that $g(r)$ in equation (\ref{eq: lambda uniform, B constant}) has a light tail.

\subsubsection{$\lambda$ is Gamma distributed}
Assume that $\lambda$ has a Gamma distribution with parameters $a$ and $b$, i.e. its probability density is $f_\lambda(s)=b^a s^{a-1} e^{-bs}/\Gamma(a)$, with $a,b>0$. Then, by using equation (\ref{eq: settling rate}) we obtain
\begin{equation}
H(t)=1-\frac{b^a}{\Gamma(a)}\int_0^\infty s^{a-1}e^{-(t+b)s}ds = 1-\left( \frac{b}{t+b}\right)^a
\end{equation} 
and equation (\ref{main formula}) gives
\begin{align} 
g(r)&=\int_0^\infty \frac{2r}{B^2 t}e^{-r^2/B^2t}\cdot \frac{ab^a}{(t+b)^{a+1}} dt \nonumber \\
&=\frac{2rab^a}{B^2}\int_0^\infty e^{-r^2 x/B^2}\frac{x^a}{(1+bx)^{a+1}}dx, \qquad (x=1/t).
\label{gamma settling}
\end{align}
By a standard result for confluent hypergeometric functions, see (\cite{Slater}) pg. 506 (13.2.8), the integral in (\ref{gamma settling}) is equal to \[\Gamma(a+1)b^{-a-1}U(a+1,1,z/b), \qquad z=r^2/B^2,\] where $\Gamma$ and $U$ denote the Gamma function and the confluent hypergeometric function of the second kind, respectively. 
Let us take, as an example, the case where the parameters for  the distribution of $\lambda$ are equal, $a=b$. Then the mean of $\lambda$ is $\mathbb{E}(\lambda)=a/b=1$ and
\begin{equation}
g(r)=\frac{2r}{B^2}\Gamma(a+1)U\left(a+1,1,\frac{r^2}{B^2a}\right).
\label{eq: lambda gamma distributed}
\end{equation}
Figure \ref{fig: BK lambda random}(b) shows the plots of $g(r)$ for two cases: when $\lambda=1$, corresponding to equation (\ref{B and K density}) (continuous curve), and when $\lambda$ follows a Gamma distribution, corresponding to equation (\ref{eq: lambda gamma distributed}) (dashed curve). The plot strongly suggests, and this can be rigorously demonstrated, that the density in equation (\ref{eq: lambda gamma distributed}) has a heavy tail.

\subsection{Random diffusion parameter $B^2$}
Let the variability in spatial movement be introduced through the diffusion parameter, that is, in equation (\ref{eq: probability radius}), $B^2$ is assumed to have a predetermined distribution. 
The distribution  for the position $R(t)$ being in the interval $[r,r+dt)$, see equation (\ref{eq: probability radius}), becomes
\begin{align}\label{diffusion position}
P(r\leq R(t)<r+dr)&=dr\int_0^\infty P(r\leq R(t)<r+dr|B^2=s)f_{B^2}(s)ds \nonumber \\ 
&=dr\int_0^\infty \frac{2r}{st}e^{-r^2/st} f_{B^2}(s)ds,
\end{align} 
where $f_{B^2}$ is the probability density for $B^2$. Here we assume that $B^2$ has a Gamma distribution with parameters $a$ and $b$. Then the integral in (\ref{diffusion position}) becomes
\begin{align*}
\int_0^\infty \frac{2r}{st}e^{-r^2/st} f_{B^2}(s)ds&=\frac{2rb^a}{t\Gamma(a)}\int_0^\infty e^{-bs-r^2/st}s^{a-2}ds \\ 
&=\frac{4rb}{t\Gamma(a)} z^{(a-1)/2}K_{a-1}(2\sqrt{z}), \qquad (z=r^2b/t),
\end{align*}
where we used a standard characterization of modified Bessel functions, see (\cite{Gradshteyn}) pg. 368 (9).
Then we can write a general expression for the radial probability density that includes a distribution of the settling rate, generalizing equation (\ref{main formula}),  
\begin{equation}\label{eq: general form section 3}
g(r)= \frac{4r^ab^{(a+1)/2}}{\Gamma(a)}\int_0^\infty \frac{1}{t^{(a+1)/2}}K_{a-1}\left(2r\sqrt{\frac{b}{t}}\right)\cdot \mathcal{L}(sf_\lambda(s))(t)dt.
\end{equation} 

\subsection{Examples with $B^2$ Gamma distributed}
Let us assume, as in the previous Section, that $B^2$ is Gamma distributed with parameters $a$ and $b$.
\subsubsection{$\lambda$ is constant}
\begin{figure}
\includegraphics[scale=0.55]{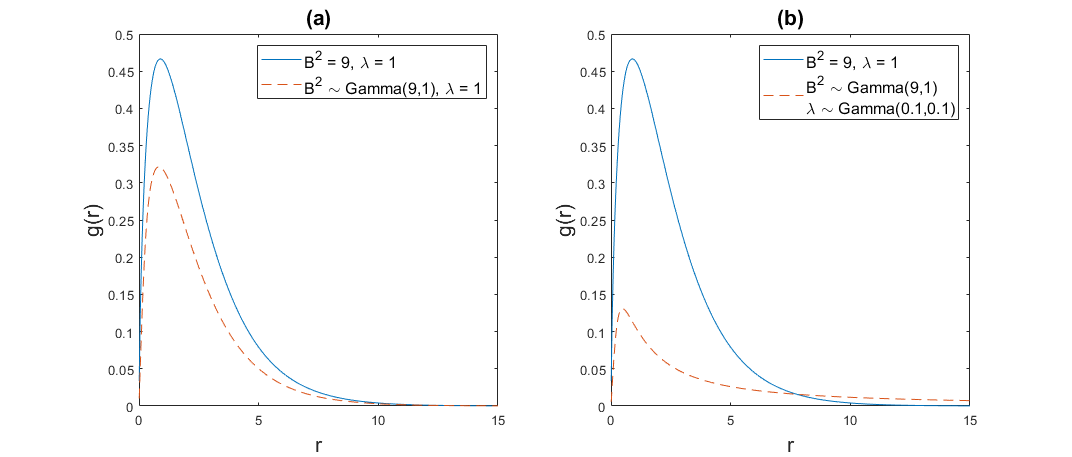}
\centering
  \caption{Probability density for the radial distribution of individuals, $g(r)$. The Broadbent-Kendall model, with $\lambda=1$ and $B^2=9$, appears as the continuous curve. For $\lambda$ (a) constant equal to 1 and (b) Gamma distributed, the densities appear as dashed curves.
  The parameters of the Gamma distributions are chosen so that  $\mathbb{E}(\lambda)=1$ and $\mathbb{E}(B^2)=9$. 
  }
\label{fig: BK both versions}
\end{figure}

If $\lambda>0$ is constant, then the radial probability density is simply
\begin{equation}
g(r)= \frac{4r^a b^{(a+1)/2}}{\Gamma(a)}\int_0^\infty \frac{1}{t^{(a+1)/2}}K_{a-1}\left(2r\sqrt{\frac{b}{t}}\right)\cdot \lambda e^{-\lambda t}dt.
\label{eq: B^2 gamma and lambda constant}
\end{equation}
In Figure \ref{fig: BK both versions}(a), we plot the probability density (\ref{eq: B^2 gamma and lambda constant}) along with the plot of the Broadbent-Kendall model (\ref{B and K density}). The parameters for the Gamma distribution of $B^2$ are chosen so that the mean of $B^2$ equals the diffusion parameter used for (\ref{B and K density}).
Although randomization causes a shift of the probability mass to the right, from Figure \ref{fig: BK both versions}(a) it appears that the resulting density remains light-tailed.

\subsubsection{$\lambda$ is Gamma distributed}
We assume now that $\lambda$ has a Gamma distribution with parameters $c$ and $d$. In this case both the diffusion and settling rate are randomized. The density for the settled organisms, equation (\ref{eq: general form section 3}), becomes
\begin{equation}
g(r)= \frac{4r^a b^{(a+1)/2}}{\Gamma(a)}\int_0^\infty \frac{1}{t^{(a+1)/2}}K_{a-1}\left(2r\sqrt{\frac{b}{t}}\right)\cdot \frac{cd^c}{(t+d)^{c+1}} dt.
\label{eq: Gamma diffusion and Gamma time}
\end{equation}
Figure \ref{fig: BK both versions}(b) shows the plots of (\ref{eq: Gamma diffusion and Gamma time}), the dashed curve, and equation (\ref{B and K density}) is the solid line. The parameter values are chosen so  that $\mathbb{E}(B^2)=9$ and $\mathbb{E}(\lambda)=1$. The shift of the probability mass to the right is significant in comparison to Figure \ref{fig: BK both versions}(a) where the tail appears to decay much more slowly. While the difference between the dashed curves in Figures \ref{fig: BK lambda random}(b) and \ref{fig: BK both versions}(b) appears negligible at first glance, a closer inspection reveals a subtle discrepancy due to the randomization of the diffusivity.

\section{Tail asymptotics}
We present the spatial tail analysis for the Broadbent-Kendall model (\ref{B and K density}), which has constant parameters $B^2$ and $\lambda$, as well as for the case where each follows a Gamma distribution, equation (\ref{eq: Gamma diffusion and Gamma time}). Similar tail analyses for the other examples in Section 2 are omitted here. The results are summarized in Table \ref{table: table 1}.

\subsection{Broadbent-Kendall model with constant diffusion and settling rate}
Let us examine the tail behavior of the density (\ref{B and K density}), derived by Broadbent and Kendall. Any modified Bessel function of the second kind has an asymptotic form, for large $z$, given by
\begin{equation}\label{Bessel asymptotics}
    K_\nu(z)\sim\sqrt{\frac{\pi}{2z}}e^{-z}\left( 1+O\left( \frac{1}{z}\right)\right),
\end{equation}
see (\cite{Nikiforov}) pg. 224 or (\cite{Olver}) pg. 378 (9.7.2). By retaining only the leading term in the expansion of \(K_0\left(2r\sqrt{\lambda}/B\right)\)
and using it in equation (\ref{B and K density}), we obtain
\begin{align*}
    g(r)\sim Cr^{1/2}e^{-kr},\qquad C=\frac{2\sqrt{\pi}\lambda^{3/4}}{B^{3/2}},\quad k=2\sqrt{\lambda}/B, 
\end{align*}
valid for large $r$. By comparing the asymptotic behavior of $g(r)$ with $e^{-\theta r}$, $\theta>0$ arbitrary, 
\begin{equation*}
    \frac{g(r)}{e^{-\theta r}}\sim C r^{1/2} e^{(\theta-k)r},
\end{equation*}
we conclude that when $\theta < k$, the ratio approaches 0 as $r \to \infty$. This means that $g(r)$ decays faster than $e^{-\theta r}$, suggesting a light tail probability. If $R$ denotes the final radial location of an individual then
\[
P(R>r) = \int_r^\infty g(s) \, ds \approx C\int_r^\infty  s^{1/2} e^{-k s} \, ds = C r^{3/2}\int_1^\infty u^{1/2}e^{-r(ku)}du \qquad(s=ru).
\]
By Laplace's method for asymptotic approximation of integrals (\cite{Murray}), the integral on the right behaves as
\[\int_1^\infty u^{1/2}e^{-r(ku)}du\sim \frac{1}{rk}e^{-kr}\]
when $r$ is large. Consequently, \(P(R>r)\sim Ck^{-1}r^{1/2} e^{-kr}\), ensuring that the tail decays exponentially.

\subsection{Broadbent-Kendall model with randomized diffusion and settling rate}
We now analyze the asymptotic behavior of $g(r)$ defined in (\ref{eq: Gamma diffusion and Gamma time}). Recall that $g(r)$ now includes simultaneous randomization of diffusion and settling parameters. The leading term of the asymptotic form of the modified Bessel function (\ref{Bessel asymptotics}) becomes
\[K_{a-1}\left(2r\sqrt{\frac{b}{t}}\right)\sim\sqrt{\frac{\pi}{4r}}\left(\frac{t}{b} \right)^{1/4} e^{-2r\sqrt{b/t}}.\]
By replacing it in (\ref{eq: Gamma diffusion and Gamma time}) we obtain
\begin{equation}
g(r)\sim\frac{2\sqrt{\pi}r^{a-1/2}b^{(2a+1)/4}cd^c}{\Gamma (a)}\int_0^\infty t^{-(2a+1)/4}e^{-2r\sqrt{b/t}}\cdot(t+d)^{-(c+1)}dt.
\label{first approximation}
\end{equation}
We use the change of variable $u=\sqrt{b/t}$ to rewrite the integral in (\ref{first approximation}) as
\begin{equation}
2b^{(3-2a)/4}\int_0^\infty u^{(2a+4c+1)/2-1}e^{-2ru}(b+du^2)^{-(c+1)}du.
\label{second approximation}
\end{equation}
The exponential $e^{-2ru}$ peaks at $u=0$, suggesting that we use Laplace's method to approximate the integral in (\ref{second approximation}). However, near $u=0$, $b+du^2\approx b$ and therefore (\ref{second approximation}) can be approximated as
\begin{equation}
2b^{(3-2a)/4-(c+1)}\int_0^\infty u^{(2a+4c+1)/2-1}e^{-2ru}du = 2b^{(3-2a)/4-(c+1)} \frac{\Gamma((2a+4c+1)/2)}{(2r)^{(2a+4c-1)/2}}.
\label{third approximation}
\end{equation}
After replacing the right hand side of (\ref{third approximation}) in (\ref{first approximation}) and simplifying, we obtain an asymptotic expression for $g(r)$,
\begin{equation}\label{eq: asymptotic B-K randomized}
    g(r)\sim Cr^{-2c-1},\qquad C=\frac{\sqrt{\pi}2^{(3-2a-4c)/2}cd^c}{b^c\Gamma(a)}\Gamma\left(\frac{2a+4c+1}{2}\right),
\end{equation}
which has power-law decay, as $c>0$, more slowly than any exponential, 
\[\frac{g(r)}{e^{-\theta r}}\sim \frac{Cr^{-2c-1}}{e^{-\theta r}}=\frac{Ce^{\theta r}}{r^{2c+1}}\rightarrow \infty \quad\text{ as }\quad r\rightarrow\infty.\]
By integrating the tail approximation in (\ref{eq: asymptotic B-K randomized}), we obtain 
$P(R>r)\sim r^{-2c}$, indicating a heavy-tailed distribution. 
 
\begin{table}
\centering

\begin{tabular}{||c || c| c| c||} 
 \hline
  & $\lambda$ constant & $\lambda\sim \text{Uniform}$ & $\lambda\sim\text{Gamma}$ \\ [0.5ex] 
 \hline
$B^2$ constant & LT & LT & HT \\ 
 \hline
 $B^2\sim\text{Gamma}$ & LT & - & HT \\[0.5ex] 
\hline
\end{tabular}
\caption{Results of tail asymptotic analysis for the probability density $g(r)$ in the examples. LT = light tail, HT = heavy tail.}
\label{table: table 1}
\end{table}

\subsection{The tale of two tails: Emergence of the heavy tail explained}
We observe in our examples that the randomization of parameters leads to heavy-tailed distributions specifically in the cases where $\lambda$ follows a Gamma distribution, as shown in Table \ref{table: table 1}. The underlying reason is that compounding a Gamma and an exponential distribution results in a shifted Pareto distribution, see equations (\ref{gamma settling}) and (\ref{eq: Gamma diffusion and Gamma time}), a heavy-tailed distribution; this is a standard result, see (\cite{Harris}) or (\cite{Johnson}) pg. 574. Thus, in the process of obtaining $g(r)$, the probability mass in equation (\ref{eq: probability radius}), whose spatial tale is light, is re-weighted in $t$ by a shifted Pareto distribution, whose temporal tale is heavy, through equation (\ref{main formula}), transfering the heavy-tailed nature of the latter to space, as was shown above.
Other distributions besides the Gamma, such as the inverse Gamma which is itself heavy-tailed, can be compounded with the exponential to yield heavy-tailed distributions for stopping times. It can be shown, following similar analysis, that the resulting probability density $g(r)$ decays more slowly than the exponential.

\section{Conclusions and discussion}
The introduction of variability in the parameters of a dispersal kernel is expected to increase its variance and affect the shape of the tail, but quantifying those changes can be challenging. 
For the case of the Broadbent-Kendall model with randomized parameters, we presented analytical expressions that allow an straightforward tail analysis.
The novelty in this paper lies in showing that the Broadbent-Kendall model, which exhibits a light-tailed distribution under fixed parameters, can generate heavy tails through parameter randomization. 
In particular, introducing randomness into the settling rate via a Gamma distribution yielded heavy-tailed behavior, with the tail probability decaying as a power law with exponent $2c$, where $c>0$ is the shape parameter of the distribution.

A basic deterministic model for dispersal and settling of propagules, analogous to the Broadbent-Kendall model, consists of a reaction-diffusion system in which individuals move via diffusion and settle at some rate $h>0$ (\cite{Lewis, Neubert, Okubo}).
In one dimension, the dispersal kernel they obtained for settled propagules is the Laplace kernel, which is light-tailed. 
By making the settling rate time dependent, $h=h(t)$, with specific functions the authors in (\cite{Neubert}) obtain new dispersal kernels (or equivalently, density functions of locations), which are also light-tailed. 
A sophisticated treatment of the reaction-diffusion equation that includes multiple scales was used in (\cite{Powell}) to explain the so called Reid's paradox (\cite{Clark98}). This paradox describes the discrepancy between theoretical and observed plant migration rates during the early Holocene period. Though first noted with oaks in Britain, it applies broadly across species for which rare, long-distance dispersal events, often facilitated by animals, remain difficult to explain. In (\cite{Powell}), it is shown by the use of spatial homogenization techniques that the spatially explicit model for active seed dispersal, with a spatial-dependent diffusion, combined with a minimal model of seed consumer foraging and caching behavior, may sufficiently explain some observed anomalous dispersal rates. The dispersal kernel obtained in (\cite{Powell}) is also a light-tailed Laplace distribution. These examples suggest that minimal deterministic dynamical models capable of producing dispersal kernels with tails that decay more slowly than exponential may be difficult to identify. However, as we have shown here, heavy-tailed dispersal kernels can arise naturally when parameter stochasticity is introduced in the settling time, even in simple models. 

Some previous efforts to explain rare, long-distance dispersal events have emphasized variability in the diffusion parameter (\cite{Clark99, Petrovskii08, Petrovskii09}), modeling it with heavy-tailed distributions, which is appropriate when large fluctuations in diffusivity occur. Those studies, however, do not include stopping times of the diffusion process. In contrast, our model includes a biological mechanism for obtaining a heavy tail.
Our complementary approach of incorporating variability in the settling rate has the potential of generating heavy-tailed distributions by exploiting the properties of compounding a Gamma with the exponential distribution for the settling times, when the randomization of the diffusivity follows a Gamma distribution and even with constant diffusivity. 

Our models suggest the need for future explorations of long distance settling patterns in space linked to the stopping times of the stochastic movement process.
\\
\\
\noindent\textbf{Data Availability Statement:}
This study does not include any data. All results are derived from theoretical analysis and mathematical modeling, and no datasets were generated or analyzed during the current study.
\bibliographystyle{plainnat}
\bibliography{main}

\end{document}